\documentclass[3p,twocolumn]{elsarticle}
\usepackage[caption = false]{subfig}
\usepackage{amsmath}
\usepackage{natbib}
\usepackage[caption = false]{subfig}
\usepackage{multirow}
\usepackage{cleveref}
\usepackage{float}
\usepackage{placeins}
\usepackage{xcolor}

\crefname{equation}{Eq.}{Eqs.}
\crefname{figure}{Fig.}{Figs.}
\crefname{table}{Table}{Tables}
\labelcrefformat{equation}{#2#1#3}

\title{Strain-Tunable Opto-electronics in PdS$_2$ Monolayer: the Role of Band Nesting and Carrier-Phonon Scattering}

\author[1,2]{Hongfa Wang}
\author[1,2]{Yancheng Gong}
\author[3]{Subrahmanyam Pattamatta}
\author[4]{Junwen Li}
\author[1,2]{Hailong Wang\corref{cor1}}
\ead{hailwang@ustc.edu.cn}
\author[3]{Zhizi Guan\corref{cor2}}
\ead{zhiziguan@connect.hku.hk}

\cortext[cor1]{Corresponding author: hailwang@ustc.edu.cn}
\cortext[cor2]{Corresponding author: zhiziguan@connect.hku.hk}

\affiliation[1]{organization={CAS Key Laboratory of Mechanical Behavior and Design
of Materials, Department of Modern Mechanics, University of Science and Technology of China},addressline={Hefei, Anhui 230027},country={China}}
\affiliation[2]{organization={State Key Laboratory of Nonlinear Mechanics, Institute of Mechanics, Chinese Academy of Science},addressline={15 Beisihuan West Road, Beijing 100190},country={China}}
\affiliation[3]{organization={Department of Mechanical Engineering, The University of Hong Kong},city={Hong Kong},country={China}}
\affiliation[4]{organization={DFTWorks LLC},city={Oakton, VA},postcode={22124},country={USA}}

\begin{document}

\begin{frontmatter}

\begin{abstract}
        Strain engineering is a powerful strategy for tuning the optoelectronic properties in two-dimensional materials, yet the underlying mechanisms governing their strain response are often not fully elucidated. In this work, our first-principle calculations show that the penta-orthorhombic PdS$_2$ monolayer exhibits two key strain-tunable properties: a continuous redshift of its main optical absorption peak from $\sim$2.0 to $\sim$1.6~eV and enhancement in carrier mobility, with a more than threefold increase for electron under 0--4\% biaxial tensile strain. Subsequent analysis reveals that the tunable optical response originates from a robust band nesting feature between the highest valence and lowest conduction bands, which is preserved across the Brillouin zone under biaxial strain. For the carrier transport, deformation potential theory predicts mobility increasing with strain, strongly correlating with the reduction of carrier effective mass.  Our first-principles calculations show a strain-induced monotonic decrease in carrier linewidths near the band edges, indicating suppressed carrier-phonon scattering and longer carrier lifetime as the origin of the mobility enhancement. Our work establishes a pathway for engineering the optoelectronic response in 2D semiconductors where strong band nesting governs the optical properties and paves the way for the rational design of continuously tunable flexible optoelectronic devices.
\end{abstract}


\end{frontmatter}

\section{Introduction}

The advent of two-dimensional (2D) materials, initiated by the discovery of graphene\cite{novoselov2004electric}, has unlocked new frontiers for next-generation flexible optoelectronics\cite{adma201702678,adfm202010533,acsami9b18945}. Novel 2D materials has paved the way for revolutionary application scenarios\cite{weardevice,acssensors4c00307}, and it is crucial to move beyond their pristine, static properties to a dynamic control of the material behavior. Unlike methodologies such as chemical modification\cite{cheeng} or defect engineering\cite{defeng} that permanently alter a material, strain engineering\cite{strain1,strain2} has emerged as a powerful paradigm to both actively and reversibly tune the optoelectronic properties of 2D materials within their elastic limits. For example, strain is widely used to tune fundamental properties such as the band gap, optical absorption, and carrier mobility in different 2D semiconductors\cite{MARIOGALICIAHERNANDEZ2022111144,acsami2c23163,Sun_2024,Kansara_2019,PhysRevB104205432}. This inherent tunability under mechanical deformation is precisely the principle that underpins flexible electronics, making the search for materials with a predictable and continuous strain response central to future device applications.

Among the diverse family of 2D materials, penta-orthorhombic PdS$_2$ monolayer has recently emerged as a promising candidate. While its bulk form was first synthesized in 1956\cite{first}, recent advances have demonstrated the synthesis of large-scale few-layer samples\cite{smll202206915,acsami1c11824}. Following its isolation in two-dimensional form, the fundamental properties of the PdS$_2$ monolayer have been extensively investigated\cite{Raval_Gupta_Gajjar_Ahuja_2022,acsanm8b00363}, establishing it as an indirect band gap semiconductor with theoretically predicted anisotropic carrier mobilities in the range of 40--340~$\text{cm}^2/\text{Vs}$\cite{C5TC01345C}. These promising optoelectronic characteristics have motivated the exploration of its potential applications. For instance, He et al. reported that PdS$_2$ photodetectors exhibit a broadband photoresponse covering 450--1550 nm\cite{acsami1c11824}, while Ma et al. investigated its suitability for toxic gas sensing\cite{acsanm3c02177}.
Furthermore, initial studies have explored its potential for strain engineering, revealing that its optoelectronic properties, such as the band gap and refractive index, are highly sensitive to external strain fields\cite{AHMAD2020100976,8681734,acsanm8b00363}.

While these studies confirm the strain sensitivity of the PdS$_2$ monolayer, the underlying mechanisms remain largely unexplored. Specifically, a comprehensive physical explanation are required for two key observations: the continuously tunable optical response and the strain-enhanced carrier mobility. %The latter is often attributed to variations in carrier effective mass within the deformation potential theory, a model that provides only a partial picture by overlooking the full scope of carrier-phonon interactions.
In this work, we bridge these gaps through systematic first-principles investigations. We show that the monotonic tunability of the optical absorption spectrum stems from a remarkably robust band nesting feature between the highest valence and lowest conduction bands. Moreover, going beyond the deformation potential theory, we conduct a first-principles simulation based on the analysis of electron-phonon coupling effect. This approach allows us to reveal that a strain-induced suppression of carrier scattering is the key factor driving the mobility enhancement, a finding corroborated by the calculated reduction in carrier linewidth. Our findings establish a deeper understanding of the strain response in the PdS$_2$ monolayer, offering insights into how biaxial strain can be used to engineer the properties of other 2D materials that exhibit similar band features. The insights, in turn, highlights the potential of the PdS$_2$ monolayer for developing continuously tunable optoelectronic devices.

\section{Computational Methods and Fundamental Properties}
We carry out the density functional theory calculations using the Vienna \textit{ab initio} simulation package~(VASP) with the exchange-correlation functional evaluated with the Perdew-Burke-Ernzerhof~(PBE) generalized gradient approximation \cite{kresse1993ab,kresse1996efficient,blochl1994projector}. The energy cutoff used in the plane wave basis expansion is set to be 450 eV and a vacuum space of 12~\AA\ along the direction normal to the PdS$_2$ monolayer is used to eliminate the interaction between artificial periodic monolayers.  We use a 9 $\times$ 9 $\times$ 1  $k$ grid for sampling the Brillouin zone.  All atoms are allowed to relax until the forces acting on each atom are less than 0.001~eV/\AA. The dielectric function and joint density of state (JDOS) is calculated using the Quantum ESPRESSO package\cite{Giannozzi_2009,Giannozzi_2017}. The phonon dispersion curves and carrier scattering rates are calculated using the ABINIT software package \cite{abinit2020}. 

\begin{figure}
  \includegraphics{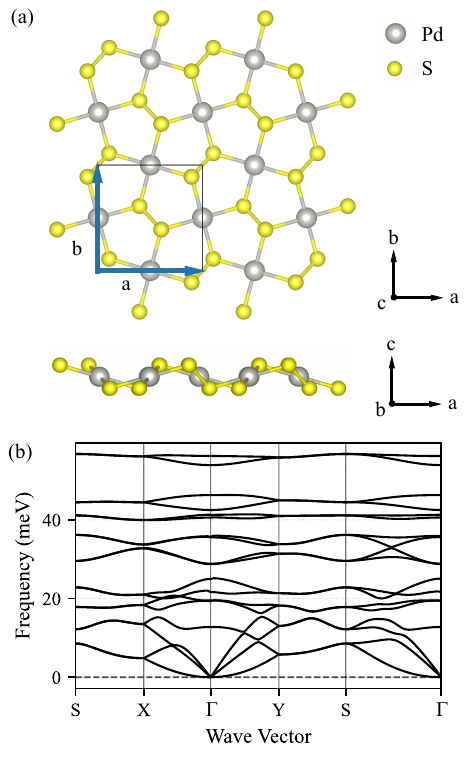}
  \centering
  \caption{\label{fig:structure} (a) The crystal structure of PdS$_2$ monolayer. Top view (upper panel) and side view (lower panel). The PdS$_2$ monolayer exhibits a penta-orthorhombic structure with lattice parameters of $a$ = 5.47 Å and $b$ = 5.57 Å. (b) Phonon dispersion of the PdS$_2$ monolayer without imaginary frequencies, confirming dynamical stability.}
\end{figure}

\begin{figure}
  \includegraphics[width=\linewidth]{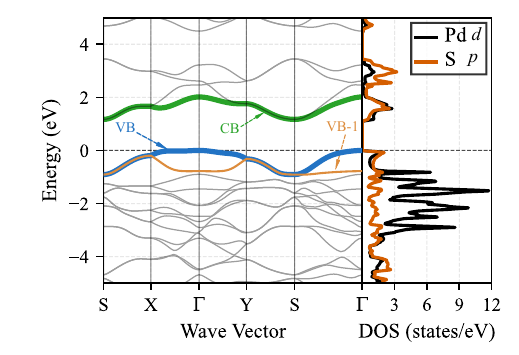}
  % Here
  \caption{\label{fig:BandDos} 
  The electronic band structure (left panel) and projected density of states (right panel) of PdS$_2$ monolayer. 
  The band structure highlights the lowest conduction band (CB, green), the highest valence band (VB, blue), and the second-highest valence band (VB-1, orange). 
  Nearly constant CB-VB separation along the $k$-paths reveals band nesting, while the projected density of states (PDOS) indicates that the states close to the VB and CB are mainly derived from the $d$ orbitals of Pd atoms (black line) and the $p$ orbitals of S atoms (red line).}
\end{figure}

The relaxed PdS$_2$ monolayer crystallizes in a penta-orthorhombic structure ($C_{2h}$ point group) with a puckered geometry reminiscent of black phosphorus (\cref{fig:structure}(a))\cite{pnas1416591112,C7RA06903K,bp}. Our first-principles calculations confirm that this structure is both mechanically and dynamically stable, with calculated elastic constants satisfying the Born-Huang criteria and a phonon dispersion free of imaginary modes (\cref{fig:structure}(b))\cite{BornHuang,LI2020100615}. Detailed data are provided in the Supplementary Information (page 1). Band structure shows the PdS$_2$ monolayer is a semiconductor with an indirect PBE-level bandgap of 1.12 eV (\cref{fig:BandDos}). Band structure is characterized by a particularly flat valence band maximum (VBM) near the $\Gamma$ point, which leads to a large hole effective mass and a corresponding sharp peak in the density of states. Orbital-resolved analysis reveals that the states near the band edges primarily arise from the $d$ orbitals of Pd atoms and the $p$ orbitals of S atoms. Furthermore, the material's non-polar nature, confirmed by the absence of LO-TO splitting in the phonon spectrum at the $\Gamma$ point (\cref{fig:structure}(b)), is a crucial factor influencing its charge transport properties. This non-polarity implies that carrier scattering is primarily dominated by acoustic phonons, a key insight for understanding the mobility results discussed later.

\begin{figure}
    \includegraphics{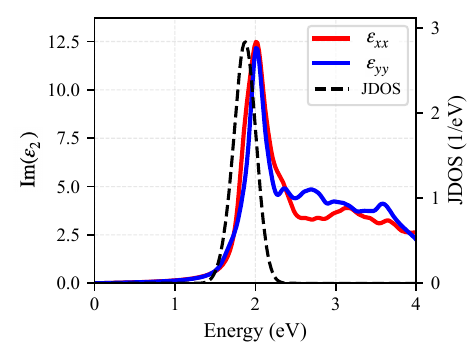}% Here is how to import EPS art
    \caption{\label{fig:epsilon}Imaginary part of the dielectric function $\mathrm{Im}(\varepsilon_{2})$ (left vertical axis) and joint density of states (JDOS, right vertical axis) between the highest valence band and lowest conduction band. }
  \end{figure}

\section{Strain-Tunable Optical Properties: The Role of Band Nesting}
% Alternative titles (commented out):
% \textcolor{blue}{\section{Strain-Tunable Optical Properties}}
% \textcolor{red}{\section{Strain-Tunable Optical Properties and Band Nesting}}
% \textcolor{green}{\section{Optical Properties, Band Nesting, and Strain Modulation}}

We find that pristine PdS$_2$ monolayer shows a strong optical absorption peak around 2.0 eV (Fig. 3), which can be continuously and monotonically tuned with external strain (see Fig. S3). To unravel the origin of this tunable optical response, we investigated the dielectric function of which the imaginary part is derived from the electronic band structure using Fermi's golden rule:
\begin{equation}\label{eq_eps}
  \varepsilon_2(\omega) = \frac{2\pi e^2}{\Omega \varepsilon_0} \sum_{v,c,\mathbf{k}} \delta(E_{c\mathbf{k}} - E_{v\mathbf{k}} - \hbar\omega) \left|\langle\psi_\mathbf{k}^c|\hat{D}|\psi_\mathbf{k}^v\rangle\right|^2
\end{equation}
where $\psi_\mathbf{k}^c$ and $\psi_\mathbf{k}^v$ denote the electronic wave functions of the conduction and valence bands at momentum $\mathbf{k}$, respectively.
$\varepsilon_0$ is the vacuum dielectric constant, and $\hat{D}$ represents the dipole transition operator.
$\hbar\omega$ is incident photon energy, $\Omega$ is the volume of the lattice cell, and $v$ and $c$ refer to the valence and conduction bands, respectively.
$E_{v\mathbf{k}}$ and $E_{c\mathbf{k}}$ are the eigenvalues of the Hamiltonian. 

\begin{figure*}[!htbp]
    \includegraphics{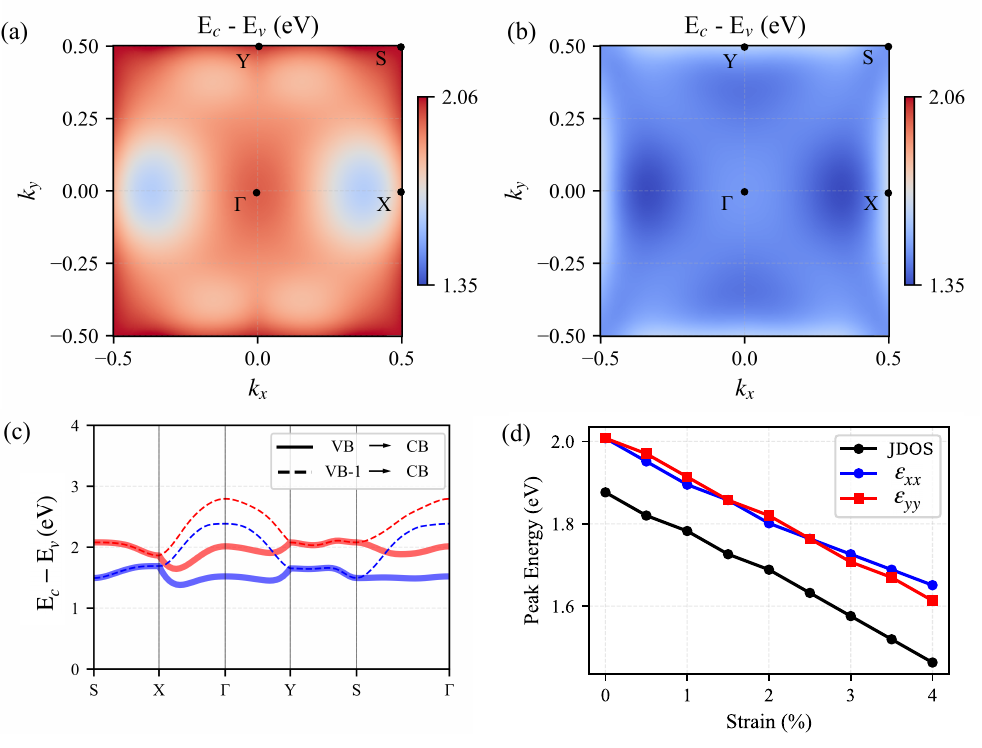}% Here is how to import EPS art and
    \caption{\label{fig:optics2} Band energy difference map between the highest valence band (VB) and lowest conduction band (CB) of PdS$_2$ monolayer across the Brillouin zone under (a) 0 and (b) 4\%, with high symmetry points in the first Brillouin zone labeled. (c) Band energy difference along high-symmetry paths for transitions from the VB (solid lines) and VB-1 (dashed lines) to the CB. Red and blue curves correspond to 0\% and 4\% strain, respectively. (d) Peak energy (horizontal axis position) of \cref{fig:epsilon} of the imaginary part of the dielectric function $\mathrm{Im}(\varepsilon_{2})$ and joint density of states (JDOS) versus strain.}
  \end{figure*}

Our analysis reveals that the prominent absorption peak at 2.0 eV can be attributed to a band nesting feature, where the highest valence band (VB) and the lowest conduction band (CB) run nearly parallel to each other across a large region of the Brillouin zone. This is validated by the band energy difference map (\cref{fig:optics2}(a)), where the energy difference between the VB and CB remains remarkably uniform across the Brillouin zone, primarily ranging from 1.6 to 2.0 eV. This uniformity creates optimal conditions for parallel interband transitions over an extended $k$-space region. Further analysis along high-symmetry paths (\cref{fig:optics2}(c)) corroborates this finding: the energy difference for the VB-to-CB transition (solid lines) is exceptionally flat around 1.9 eV. In contrast, transitions from the lower valence band (VB-1, dashed lines) show greater dispersion, indicating a weaker nesting effect. The nesting effect can be quantified by the JDOS, which shows the number of available transition states at a given photon energy. The JDOS spectrum (JDOS$(\hbar\omega)$), computed as $\sum_{v,c} \frac{\Omega}{(2\pi)^3}\int d^3\mathbf{k}\delta(E_{c\mathbf{k}}-E_{v\mathbf{k}}-\hbar\omega)$, shows a sharp, Van Hove-like peak at approximately 1.9 eV (\cref{fig:epsilon}). The excellent agreement between th JDOS peak and the main absorption peak in $\varepsilon_2(\omega)$ confirms that the strong optical response originates from the band nesting feature.

Having established band nesting as the key mechanism, we then explored its evolution under biaxial strain. As shown in \cref{fig:optics2}(d), the optical absorption peak and the JDOS peak both exhibit a continuous, monotonic redshift with applied tensile strain. Specifically, under 4\% strain, the dielectric function peak shifts from $\sim$2.0 eV to $\sim$1.6 eV, with the JDOS peak shifting from $\sim$1.9 eV to $\sim$1.5 eV. This behavior is a direct consequence of the robust preservation of the band nesting feature. The strained energy difference map in \cref{fig:optics2}(b) remains remarkably uniform, similar to the unstrained case, though the entire energy landscape is uniformly redshifted. Similarly, the energy difference curves in \cref{fig:optics2}(c) show that the strained (blue) and unstrained (red) profiles retain nearly identical shapes, merely shifted to lower energies. Such preservation of parallel bands is also relevant to the formation of excitons, analogous to the well-known C-exciton in monolayer MoS$_2$\cite{Sun2022,acsjpcc7b07939,PhysRevB88115205,Li106350060587}. In PdS$_2$ monolayer, the robust and tunable excitonic response under strain provides a direct pathway toward developing advanced optoelectronic devices, such as dynamically tunable photodetectors, wide-range optical modulators, and sensitive strain sensors that translate mechanical inputs into measurable optical signals\cite{PhysRevB110245310,adma202205714,Kianinia106350072091,zhao0864}.

%For example, sun\cite{Sun2022}, zhao\cite{acsjpcc7b07939}, Carvalho\cite{PhysRevB88115205}

\section{Strain-Enhanced Carrier Mobility: the Role of Carrier-Phonon Scattering}
% Alternative titles (commented out):
% \textcolor{blue}{\section{Mechanism of Strain-Enhanced Carrier Mobility}}
% \textcolor{green}{\section{Strain Modulation of Mobility and Carrier-Phonon Scattering}}

Carrier mobility is a crucial parameter that describes the movement of electrons or holes in response to an applied electric field. It characterizes the charge transport properties and is essential for evaluating the electronic transport performance of different 2D materials. The carrier mobility is first determined using the deformation potential~(DP) theory derived by Bardeen and Shockley under the effective mass approximation and the carrier-acoustic phonon scattering mechanism\cite{bardeen1950deformation}:

\begin{equation} \label{eq1}
  m^{*} = \hbar^2 (\frac{d^2 E_k}{dk^2})^{-1}
  \end{equation}
  
  \begin{equation} \label{eq2}
  \mu_{2D} = \frac{e \hbar^{3} C_{2d}}{k_{B} T |m^{*} m_{d} | (E_1)^{2}}
  \end{equation}

where $k_B$ is the Boltzmann constant, $e$ is the electron charge, $\hbar$ is the reduced Planck constant, $m^{*}$ is the carrier effective mass corresponding to the transport direction, $m_d = \sqrt{m^{*}_a m^{*}_b}$  is the average carrier effective mass, $C_{2d}$ is the elastic constant and $E_1$ is the deformation potential constant.  

\begin{figure*}[!htbp]
    \includegraphics{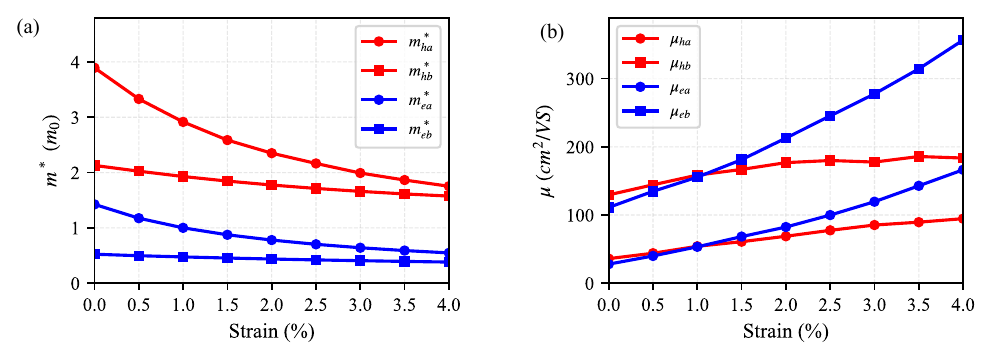}
    \caption{\label{fig:mobility} Plotted for both the $a$ and $b$ directions are: (a) carrier effective mass, and (b) the resulting carrier mobility as a function of biaxial tensile strain. The elastic constant and deformation potential under biaxial tensile strain are shown in Fig. S5 and Fig. S6, respectively.}
  \end{figure*}

Based on this framework, our calculations for the unstrained PdS$_2$ monolayer yield electron mobilities of 28 and 111 $\text{cm}^2/\text{Vs}$ and hole mobilities of 36 and 130 $\text{cm}^2/\text{Vs}$ along the $a$ and $b$ directions, respectively (\cref{table:mobility}). Upon applying biaxial tensile strain, a significant enhancement in mobility is observed for both carrier types, as shown in \cref{fig:mobility}(d). For electrons, the mobility enhancement is driven by a monotonic decrease in both effective mass and deformation potential. For holes, the situation involves competing factors, where the mobility increase is dominated by the reduction in effective mass, which outweighs the opposing contributions from the other parameters (see Fig. S5 and Fig. S6 for elastic constant and deformation potential under strain). Thus, within the DP model, the enhancement of mobility for both carrier types is primarily linked to the reduction of their effective mass under strain.

\begingroup
\setlength{\tabcolsep}{5pt} % 减小列间距
\renewcommand{\arraystretch}{1.125} 
\begin{table*}[!htbp]  % 改为单栏table
\centering
\small  % 使用小字体
\caption{Elastic Constant $C_{2D}$, Deformation Potential Constant $E_1$, Effective Mass $m^*$, and Carrier Mobility $\mu$}
\begin{tabular}{ccccccccccccc}\hline
  \multirow{2}{*}{strain (\%)} & \multirow{2}{*}{carrier type} & \multicolumn{2}{c}{$C_{2D}~(\rm{N}~\rm{m}^{-1})$} & & \multicolumn{2}{c}{$E_1~(\rm{eV})$}& & \multicolumn{2}{c}{$m^*~($m$_0)$} && \multicolumn{2}{c}{$\mu~(\times 10^2 \rm{cm}^2/\rm{Vs})$} \\  \cline{3-4} \cline{6-7} \cline{9-10} \cline{12-13}
  &  & $C_{11}$ & $C_{22}$ & & $a$ & $b$ & & $a$ & $b$ && $a$ & $b$ \\ \hline
  
  \multirow{2}{*}{0} & $e$ & \multirow{2}{*}{57.42} & \multirow{2}{*}{79.46} &  & 5.96 & 5.80 & & 1.43 & 0.52 && 0.28 & 1.11  \\
  & $h$ &  &  &   & 1.74 & 1.46 & & 3.89 & 2.13 && 0.36 & 1.30 \\ 

  \multirow{2}{*}{0.5} & $e$ & \multirow{2}{*}{57.82} & \multirow{2}{*}{78.03} &  & 5.86 & 5.70 & & 1.18 & 0.50 && 0.40 & 1.35  \\
  & $h$ &  &  &   & 1.80 & 1.48 & & 3.33 & 2.03 && 0.44 & 1.44 \\   

  \multirow{2}{*}{1} & $e$ & \multirow{2}{*}{58.19} & \multirow{2}{*}{76.60} & & 5.80 & 5.66 && 1.00 & 0.48 && 0.53 & 1.55 \\
  & $h$ &  &  & &  1.82 & 1.50 && 2.92 & 1.93 && 0.54 & 1.58 \\ 

  \multirow{2}{*}{1.5} & $e$ & \multirow{2}{*}{58.33} & \multirow{2}{*}{74.93} & & 5.74 & 5.54 && 0.88 & 0.46 && 0.68 & 1.81 \\
  & $h$ &  &  & &  1.90 & 1.54 && 2.59 & 1.85 && 0.61 & 1.67 \\  

  \multirow{2}{*}{2} & $e$ & \multirow{2}{*}{58.42} & \multirow{2}{*}{73.25} & & 5.76 & 5.36 && 0.78 & 0.44 && 0.82 & 2.13 \\
  & $h$ &  &  & &  1.94 & 1.56 && 2.35 & 1.78 && 0.69 & 1.77  \\
  
  \multirow{2}{*}{2.5} & $e$ & \multirow{2}{*}{58.29} & \multirow{2}{*}{71.34} & & 5.70 & 5.20 && 0.70 & 0.42 && 0.99 & 2.45 \\
  & $h$ &  &  & &  1.96 & 1.60 && 2.17 & 1.71 && 0.78 & 1.80  \\

  \multirow{2}{*}{3} & $e$ & \multirow{2}{*}{58.02} & \multirow{2}{*}{69.36} & & 5.62 & 5.06 && 0.64 & 0.41 && 1.19 & 2.78 \\
  & $h$ &  &  & &  2.00 & 1.66 && 1.99 & 1.66 && 0.85 & 1.78  \\
  
  \multirow{2}{*}{3.5} & $e$ & \multirow{2}{*}{57.69} & \multirow{2}{*}{67.31} & & 5.50 & 4.90 && 0.59 & 0.39 && 1.43 & 3.14 \\
  & $h$ &  &  & &  2.06 & 1.66 && 1.87 & 1.61 && 0.89 & 1.86  \\

  \multirow{2}{*}{4} & $e$ & \multirow{2}{*}{57.08} & \multirow{2}{*}{65.17} & & 5.40 & 4.72 && 0.55 & 0.38 && 1.66 & 3.56 \\
  & $h$ &  &  & &  2.10 & 1.70 && 1.75 & 1.58 && 0.95 & 1.84  \\
  \hline
\end{tabular}
\label{table:mobility}
\end{table*}
\endgroup

However, the DP approximation provides an incomplete picture by considering only long-wavelength acoustic phonons. A complete understanding requires accounting for the total scattering rate from all phonon modes, which is fundamentally linked to carrier mobility ($\mu$). The total scattering rate determines the carrier lifetime ($\tau$), the average time between scattering events, and in transport theory, mobility is directly proportional to this lifetime ($\mu \propto \tau$). The lifetime, in turn, is related to the energy broadening of the carrier state, i.e., the linewidth ($\Gamma$), by the uncertainty principle, where $\Gamma \propto 1/\tau$. Combining these relationships yields an inverse proportionality between mobility and linewidth: $\mu \propto 1/\Gamma$. Therefore, a narrower linewidth directly signifies weaker overall scattering and consequently, higher carrier mobility, making the first-principles calculation of the carrier linewidth a powerful method for investigating the fundamental origins of mobility\cite{RevModPhys.89.015003,PhysRevB.110.L121106,PhysRevB.101.165102}. We performed such calculations within the framework of density functional perturbation theory and the self-energy relaxation time approximation (SERTA)\cite{Ponc_2020,PhysRevB102094308}, where the linewidth is given by:

\begin{equation}\label{eq_linewidth}
\begin{split}
\Gamma_{n\mathbf{k}} &= \pi \frac{1}{\Omega_{BZ}} \int_{BZ} d\mathbf{q} \sum_{m,\nu} \left| g_{mn\nu}(\mathbf{k}, \mathbf{q}) \right|^2 \\
&\quad \times \left[ (n_{\mathbf{q}\nu} + 1 - f_{\mathbf{k}+\mathbf{q}}) \delta(\varepsilon_{n\mathbf{k}} - \varepsilon_{m\mathbf{k}+\mathbf{q}} - \hbar \omega_{\mathbf{q}\nu}) \right. \\
&\quad\quad \left. + (n_{\mathbf{q}\nu} + f_{\mathbf{k}+\mathbf{q}}) \delta(\varepsilon_{n\mathbf{k}} - \varepsilon_{m\mathbf{k}+\mathbf{q}} + \hbar \omega_{\mathbf{q}\nu}) \right]
\end{split}
\end{equation}
where $g_{mn\nu}(\mathbf{k}, \mathbf{q})$ is the electron-phonon coupling matrix element, $n_{\mathbf{q}\nu}$ and $f_{\mathbf{k}+\mathbf{q}}$ are the Bose-Einstein and Fermi-Dirac distributions, and $\omega_{\mathbf{q}\nu}$ is the phonon frequency. We calculated the linewidth for states within a 100 meV energy window around the band edges, ensuring that all thermally accessible carrier states at room temperature ($k_B T \approx 26$~meV) are included.

\begin{figure}
  \includegraphics{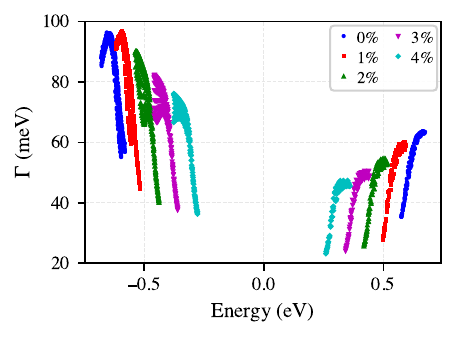}% Here is how to import EPS art
  \caption{\label{fig:linewidth} Carrier linewidth under different strain conditions, with Fermi level set to 0 eV. The energy window of calculation is 100 meV above the CBM (and 100 meV below the VBM).}
\end{figure}

The calculated carrier linewidths are presented in \cref{fig:linewidth}. The figure reveals two key features. First, for any given strain, the linewidth increases sharply as carrier energy moves away from the band edges, a characteristic behavior in 2D semiconductors where a larger phase space for scattering becomes available for higher-energy carriers \cite{SU2020147341}. Second, and more importantly, there is a clear and monotonic trend with strain: as the applied tensile strain increases from 0\% to 4\%, the linewidths for both electrons and holes at any given energy consistently decrease. Our results provide a direct explanation for the mobility enhancement. As established, carrier mobility is inversely proportional to the linewidth ($\mu \propto 1/\Gamma$). Therefore, the strain-induced reduction in linewidth shown in \cref{fig:linewidth} is the direct physical origin of the enhanced carrier mobility. This finding, where strain fundamentally alters the scattering dynamics, is consistent with recent state-of-the-art calculations\cite{PhysRevLett125177701}. For example, Ma et al. demonstrated that in monolayer $\alpha$-Te, tensile strain enhances electron mobility by suppressing the scattering rate\cite{acsami0c10236}.

Our work provides a fundamental contribution to the field of strain engineering by establishing a clear physical pathway from mechanical strain to charge transport properties. While the DP theory correctly identifies the reduction in effective mass as a key factor, our  analysis reveals a deeper mechanism: the strain-induced weakening of the overall carrier-phonon interaction, as reflected from the reduced carrier linewidth, is the primary origin of the mobility enhancement. The ability to actively suppress scattering and enhance carrier mobility is a crucial requirement for next-generation electronics\cite{Fiori2014,Schwierz2010}. Our findings pave the way for utilizing strain engineering to design high-performance flexible devices, such as high-frequency transistors and fast-response sensors, where superior charge transport is a critical performance metric\cite{Akinwande2014,10.10631.5053795}.

\section{Conclusion}

In summary, we have elucidated the underlying mechanisms of strain-tuning optoelectronic properties in PdS$_2$ monolayer through systematic first-principles investigations. We found that the continuous redshift of the main optical absorption peak is a direct consequence of a highly stable band nesting feature between the highest valence and lowest conduction bands, which persists under 4\% biaxial strain. For the transport property, our work provides a comprehensive explanation for the observed mobility enhancement. Analysis within the deformation potential approximation points to a reduction in carrier effective mass, and our first-principles electron-phonon coupling calculations complete this picture. We show that the overall carrier-phonon scattering is suppressed by strain, a finding quantified by a narrowing of the carrier linewidth. These findings provide a more complete picture of the strain response in PdS$_2$ monolayer. Our work not only confirms PdS$_2$ monolayer as a strong candidate for next-generation tunable optoelectronic and flexible electronic devices but also paves the way for a mechanism-driven approach to designing and engineering novel functionalities in other 2D materials via strain engineering.

\section{Acknowledgement}
H.W., Y.G., and H.W. acknowledge support from the National Natural Science Foundation of China (Grant No. 12172347 and No. 12232016). Z.G. also acknowledges the computational resource provided by National Supercomputer Center in Guangzhou Nansha Sub-Center.

\bibliographystyle{elsarticle-num-names} 
\bibliography{ref}

\end{document}